\documentclass{andp2012}
\usepackage[english]{babel}
\usepackage{bm}
\usepackage{color}
\usepackage{widetext}
\usepackage{caption}
\usepackage{subcaption}
\usepackage{graphicx}
\DeclareMathAlphabet\mathbfcal{OMS}{cmsy}{b}{n}

\title{Excitation of forbidden electronic transitions in atoms by Hermite-Gaussian modes}

\author[A.\,A. Peshkov]{Anton A. Peshkov\inst{1, 2}\footnote{Corresponding author\quad E-mail:~\textsf{anton.peshkov.ext@ptb.de}}}
\author[J.\,E. Jordan]{Elena Jordan \inst{1}}
\author[M. Kromrey]{Markus Kromrey \inst{1}}
\author[K. Mehta]{Karan K. Mehta \inst{3}}
\author[T.\,E. Mehlstaubler]{Tanja E. Mehlst\"a{}ubler \inst{1, 4, 5}}
\author[A. Surzhykov]{Andrey Surzhykov\inst{1, 2, 6}}

\address[1]{Physikalisch-Technische Bundesanstalt, D-38116 Braunschweig, Germany}
\address[2]{Institut f\"ur Mathematische Physik, Technische Universit\"at Braunschweig, D-38106 Braunschweig, Germany}
\address[3]{School of Electrical and Computer Engineering, Cornell University, Ithaca, NY 14850, USA}
\address[4]{Institut f\"ur Quantenoptik, Leibniz Universit\"at Hannover, D-30167 Hanover, Germany}
\address[5]{Laboratorium f\"ur Nano- und Quantenengineering, Leibniz Universit\"at Hannover, Schneiderberg 39, 30167 Hannover, Germany}
\address[6]{Laboratory for Emerging Nanometrology Braunschweig, D-38106 Braunschweig, Germany}

\shortauthors{A. A. Peshkov et al.}

\begin{abstract}
Photoexcitation of trapped ions by Hermite-Gaussian (HG) modes from guided beam structures is proposed and investigated theoretically. In particular, simple analytical expressions for the Rabi frequencies of induced atomic transitions are derived that depend both on the parameters of HG beams and on the geometry of an experiment. By using these general expressions, we investigate the $^{2}S_{1/2} \to \; ^{2}F_{7/2}$  electric octupole (E3) transition in an Yb$^{+}$ ion, localized in the low--intensity center of the HG$_{10}$ and HG$_{01}$ beams. We show how the corresponding Rabi frequency can be enhanced by properly choosing the polarization of incident light and the orientation of an external magnetic field, which defines the quantization axis of a target ion. The calculations, performed for experimentally feasible beam parameters, indicate that the achieved Rabi frequencies can be comparable or even higher than those observed for the conventional Laguerre-Gaussian (LG) modes. Since HG-like modes can be relatively straightforwardly generated with high purity and stability from integrated photonics, our results suggest that they may form a novel tool for investigating highly-forbidden atomic transitions. 
\end{abstract}
\shortabstract

\begin{document}
\maketitle

%
%
\section{Introduction}

In the past years there has been increasing interest in the use of structured light fields in microscopy \cite{Heintzmann/NP:2009}, optical tweezers \cite{Padgett/NP:2011}, manipulating cold quantum gases \cite{Andersen/PRL:2006}, as well as classical and quantum communication systems \cite{Wang/NP:2012, Krenn/PNAS:2014}. This interest is spurred by the unique properties of structured light such as spatially dependent amplitude, phase and polarization \cite{Rubinsztein-Dunlop/JO:2017, Shen/LSA:2019}. From a metrological perspective, structured light is of particular importance because of its potential to excite narrow-line clock transitions with simultaneous suppression of the undesirable light shift. This can be achieved since structured light beams exhibit highly inhomogeneous intensity profiles with a dark (low--intensity) center. An atom or ion, placed in this center, is exposed to an electric field of low strength but high gradient, which favours non--dipole transitions but weakly perturbs atomic levels. Such an ``excitation in the darkness'' has been successfully demonstrated in experiments with a single trapped $^{40}$Ca$^+$ \cite{Schmiegelow/NC:2016} and an $^{171}$Yb$^+$ ion \cite{Lange/PRL:2022}. In these experiments, ions interacted with conventional Laguerre-Gaussian (LG) modes and the excitation probability exhibited a strong dependence on both polarization and applied magnetic field orientation \cite{Solyanik-Gorgone/JOSAB:2019, Schulz/PRA:2020}. Similar studies have been performed with optical standing waves \cite{Mundt/PRL:2002, Vasquez/PRL:2023}.

Even though both experiments with $^{40}$Ca$^+$ and $^{171}$Yb$^+$ ions have provided important steps towards operating the dipole-forbidden transitions with structured light, they suffered from imprecise localization of a target ion with respect to the low-intensity laser beam center. This uncertainty in target localization arises not only from the thermal motion of an atom in a trap, but also from the lack of laser beam pointing stability. The latter problem can be partially solved by using the integrated schemes in which the light delivery optics is fabricated directly into the atom traps \cite{Mehta/NN:2016, Mehta/N:2020, Niffenegger/N:2020}.  Such photonic integrated circuits are produced using lithographic techniques that result in rectangular waveguide structures. The geometry of these structures allows fairly straightforward excitation of free-space beams approximating Hermite-Gaussian (HG) modes, which in contrast to Laguerre-Gaussian ones do not possess the rotational symmetry. In this contribution, therefore, we propose to use HG modes in photonic integrated setups for precision spectroscopy on strongly forbidden (atomic clock) transitions. To assess the feasibility of this approach, we investigate here how a single trapped Yb$^{+}$ ion interacts with Hermite-Gaussian light.

The paper is organized as follows. In Sec.~\ref{sec:theory} we briefly discuss the derivation of amplitudes for the excitation of a single trapped atom by polarized HG beams. Based on the obtained expressions we explore in Sec.~\ref{sec:results} the laser--induced $^{2}S_{1/2}  (F = 0) \to \; ^{2}F_{7/2}  (F = 3, M = 0)$  electric octupole (E3) transition in $^{171}$Yb$^{+}$ ion localized in the dark center of the beam. Special attention is paid to the question of how the Rabi frequency $\Omega^{\rm (HG)}$ of this transition is affected by the polarization of incident Hermite-Gaussian light and the orientation of an external magnetic field. Moreover, we present the results of calculations of $\Omega^{\rm (HG)}$ for realistic experimental parameters which indicate that HG modes can serve as a valuable tool for studying dipole--forbidden transitions in an integrated optical setup. Finally, Sec.~\ref{sec:summary} provides a conclusion and an outlook.

%
%
%
%
\section{Theory}
\label{sec:theory}
\subsection{Bessel modes}
\label{sec:bes}
Before we analyze the process of photoexcitation by HG modes, we first recall the simpler problem of an atom interacting with Bessel light. Since this topic has been extensively discussed in the literature \cite{Schulz/PRA:2020, Surzhykov/PRA:2015, Babiker/JO:2019, Knyazev/PU:2018}, we restrict ourselves here to basic expressions. The Bessel beam is characterized by a well-defined projection $m_\gamma$ of the total angular momentum upon its propagation direction, the helicity $\lambda$, and the longitudinal component $k_z$ of the linear momentum. Moreover, the absolute value of the transverse momentum $|\bm{k}_\perp| = \varkappa$, and hence the frequency $\omega = ck = c \sqrt{k_z^2 + \varkappa^2}$, are also fixed. The vector potential of such a Bessel beam
\begin{align}
	\label{eq:bes}
	\bm{A}^{(\text{B})}_{m_{\gamma} \lambda}  (\bm{r}; \, \varkappa, k_z) = \int a_{\varkappa m_{\gamma}} (\bm{k}_{\perp}) \, \bm{e}_{\bm{k} \lambda} \, e^{i \bm{k} \bm{r}} \, \frac{d^2 \bm{k}_{\perp}}{(2 \pi)^{2}}   \, 
\end{align}
can be written as a superposition of plane waves with wave vectors $\bm{k}$ uniformly distributed upon the surface of a cone, whose axis coincides with the light propagation direction and whose opening angle is given by $\theta_k = \arcsin (\varkappa / k)$. In Eq.~\eqref{eq:bes}, $\bm{e}_{\bm{k} \lambda}$ is the photon polarization vector satisfying the Coulomb gauge condition $\bm{e}_{\bm{k} \lambda} \cdot \bm{k} = 0$, and the amplitude $a_{\varkappa m_{\gamma}} (\bm{k}_{\perp})$ is defined by:
\begin{align}
	\label{eq:bes_amp}
	a_{\varkappa m_{\gamma}} (\bm{k}_{\perp}) = \frac{2 \pi}{\varkappa} \, (-i)^{m_\gamma} \, e^{i m_\gamma \phi_k} \, \delta (k_{\perp} - \varkappa )  \, ,
\end{align}
where $\phi_k$ is the azimuthal angle of the wave vector.
	
\begin{figure}[t!]
    \centering
	\includegraphics[width=0.96\linewidth]{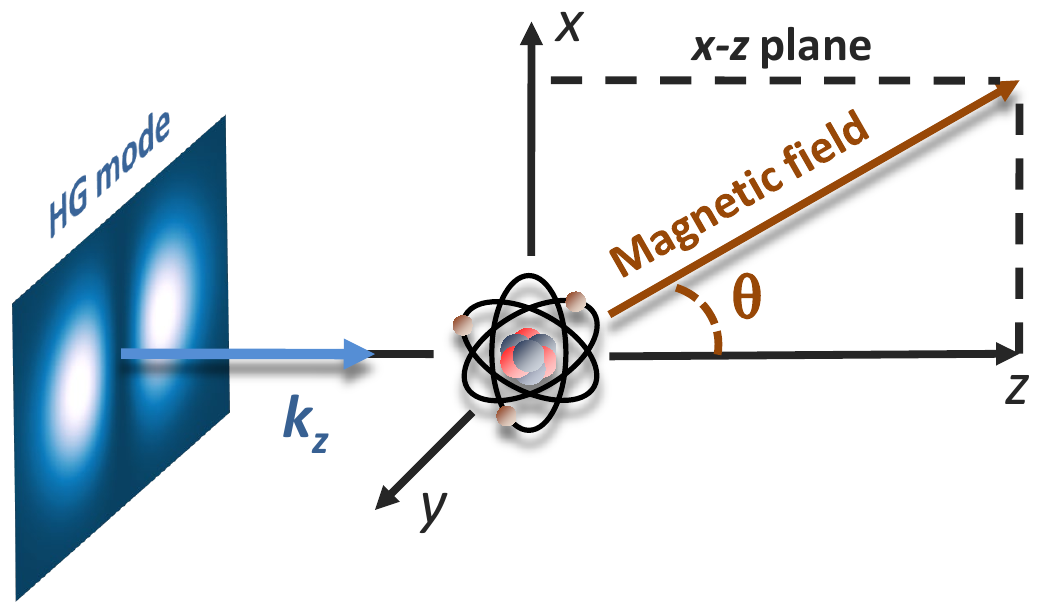}
	\caption{Geometry of the excitation of a single trapped atom by Hermite–Gaussian modes. The angle $\theta$ defines the direction of the quantization axis (determined by the external magnetic field) with respect to the light propagation direction. The $x$-$z$ plane is spanned by the light propagation axis and the magnetic field. The atom is localized in the vicinity of the beam center located at $x=y=0$.}
	\label{fig:geometry}
\end{figure}

With the help of the vector potential \eqref{eq:bes}, we can write down the amplitude for a radiative transition $|\alpha_g F_g \, M_g \rangle + \gamma \to |\alpha_e F_e \, M_e \rangle$. The ground and excited states of an atom are characterized here by nuclear $I$, electron $J$, and total $\bm{F} = \bm{I} + \bm{J}$ angular momenta, and the projection $M = M_F$ on the atomic quantization axis. This axis is directed along the external magnetic field $\bm{B}$, tilted at an angle $\theta$ with respect to the light propagation direction (see Fig.~\ref{fig:geometry}). Moreover, $\alpha$ denotes all the additional quantum numbers that are needed for a unique specification of the states. The transition amplitude is:
\begin{align}
	\label{eq:melement1}
	\mathcal{M}_{M_e M_g}^{(\text{B})} &= \Bigg  \langle \alpha_{e} F_{e} M_{e} \Bigg| \sum_{q} \bm{\alpha}_{q} \, \bm{A}_{m_{\gamma} \lambda}^{(\text{B})}  (\bm{r}_q; \, \varkappa, k_z) \Bigg| \alpha_{g} F_{g} M_{g} \Bigg \rangle \, \notag \\
	&= \int a_{\varkappa m_{\gamma}} (\bm{k}_{\perp}) \, e^{-i \bm{k}_\perp \bm{b}} \, \notag \\
	& \times \Bigg  \langle \alpha_{e} F_{e} M_{e} \Bigg| \sum_{q} \bm{\alpha}_{q} \, \bm{e}_{\bm{k} \lambda} \, e^{i \bm{k} \bm{r}_q} \Bigg| \alpha_{g} F_{g} M_{g} \Bigg \rangle \, \frac{d^2 \bm{k}_{\perp}}{(2 \pi)^{2}} \, , 
\end{align}
where $\bm{\alpha}_{q}$ is the vector of Dirac matrices for the $q$th electron, and the impact parameter $\bm{b}$ specifies the lateral position of an atom with respect to the beam axis. This parameter is introduced since Bessel and Hermite-Gaussian modes have a complex position-dependent structure. For Bessel beams, the case of zero impact parameter, $b=0$, corresponds to an atom located on a vortex line \cite{Schulz/PRA:2020}.

In order to compute the amplitude \eqref{eq:melement1}, we need to choose the quantization axis of the entire system ``atom \textit{plus} light''. The two most obvious choices are the light propagation axis and the magnetic field direction. Of course, physical observables are  independent of a particular choice of coordinate system. For the purposes of analysis, however, it is practical to take the overall quantization axis along the $\bm{B}$-field. Having defined the geometry, we can now apply the standard multipole decomposition technique to the vector potential of the radiation field \cite{Rose:1957}. Namely, the expansion of the plane-wave components of $\bm{A}^{(\text{B})}_{m_{\gamma} \lambda}$ is:
\begin{align}
	\label{eq:pw}
	\bm{e}_{\bm{k} \lambda} \, e^{i \bm{k} \bm{r}} =& \sqrt{2\pi} \sum_{pLM M'} i^L \, [L]^{1/2} \, (i \lambda)^{p} \, D^{L}_{M \lambda} (\phi_k, \, \theta_k, 0) \, \notag \\
	&\times D^{L}_{M' M} (\pi, \, \theta, \, \pi) \, \bm{a}^{(p)}_{LM'} (\bm{r}) \, , 
\end{align}
where $D^{L}_{M \lambda}$ is the Wigner $D$-function, $[L] = 2L + 1$, and $\bm{a}^{(p)}_{LM'}$ refers to magnetic $(p = 0)$ and electric $(p = 1)$ multipole potentials. In Eq.~\eqref{eq:pw}, $\theta_k$ and $\phi_k$ are angles that determine the propagation direction of each plane-wave component with respect to the beam axis and the $x$-$z$ plane, respectively. Making use of the Wigner-Eckart theorem and integrating over $\bm{k}_{\perp}$, we find:
\begin{align}
	\label{eq:melement2}
	&\mathcal{M}_{M_e M_g}^{(\text{B})} (\bm{b}) = \sqrt{2\pi} \sum_{pLM} i^L \, [L, F_g]^{1/2} \, (i \lambda)^p \, (-1)^{m_{\gamma}} \, i^M \, \notag \\
	&\;\;\; \times e^{i(m_{\gamma} - M) \phi_b} \, d^{L}_{M, \lambda} (\theta_k) \, d^{L}_{M, M_e - M_g} (\theta) \, J_{m_{\gamma} - M} (\varkappa b) \, \notag \\
	&\;\;\; \times \langle F_g M_g \, L M_e - M_g | F_e M_e \rangle \, (-1)^{J_e + I + F_g + L} \, \notag \\
    &\;\;\; \times
    \left\{ \begin{array}{ccc}
    F_e & F_g & L \\
    J_g & J_e & I
    \end{array} \right\}
    \, \langle \alpha_e J_e || H_{\gamma} (pL) || \alpha_g J_g \rangle \, , 
	\end{align}
with $d^{L}_{M, \lambda}$ and $J_{m_{\gamma} - M}$ being the small Wigner and Bessel functions, respectively. Here we have introduced, moreover, the reduced matrix element of the transition operator $ H_{\gamma} (pL) = \sum_{q} \bm{\alpha}_{q} \bm{a}^{(p)}_{L} (\bm{r}_q) $ that does not depend on the projections of the angular momenta of an atom and photon. As seen from Eq.~\eqref{eq:melement2}, the matrix element $\mathcal{M}_{M_e M_g}^{(\text{B})}$ depends on the tilt angle $\theta$ of the magnetic field, the atom's impact parameter $\bm{b}$, the opening angle $\theta_k$, the helicity $\lambda$, and the total angular momentum projection $m_{\gamma}$ of the beam. In the past, Eq.~\eqref{eq:melement2} has been employed to study transitions in atoms and ions exposed to Bessel and Laguerre-Gaussian beams \cite{Lange/PRL:2022, Afanasev/NJP:2018}. The application of this formula to the latter case is justified when an atom is located near a vortex line, where the phase and intensity patterns of LG and Bessel modes are almost identical \cite{Baghdasaryan/PRA:2019}.
\begin{figure*}[t!]
    \centering
	\includegraphics[width=0.76\linewidth]{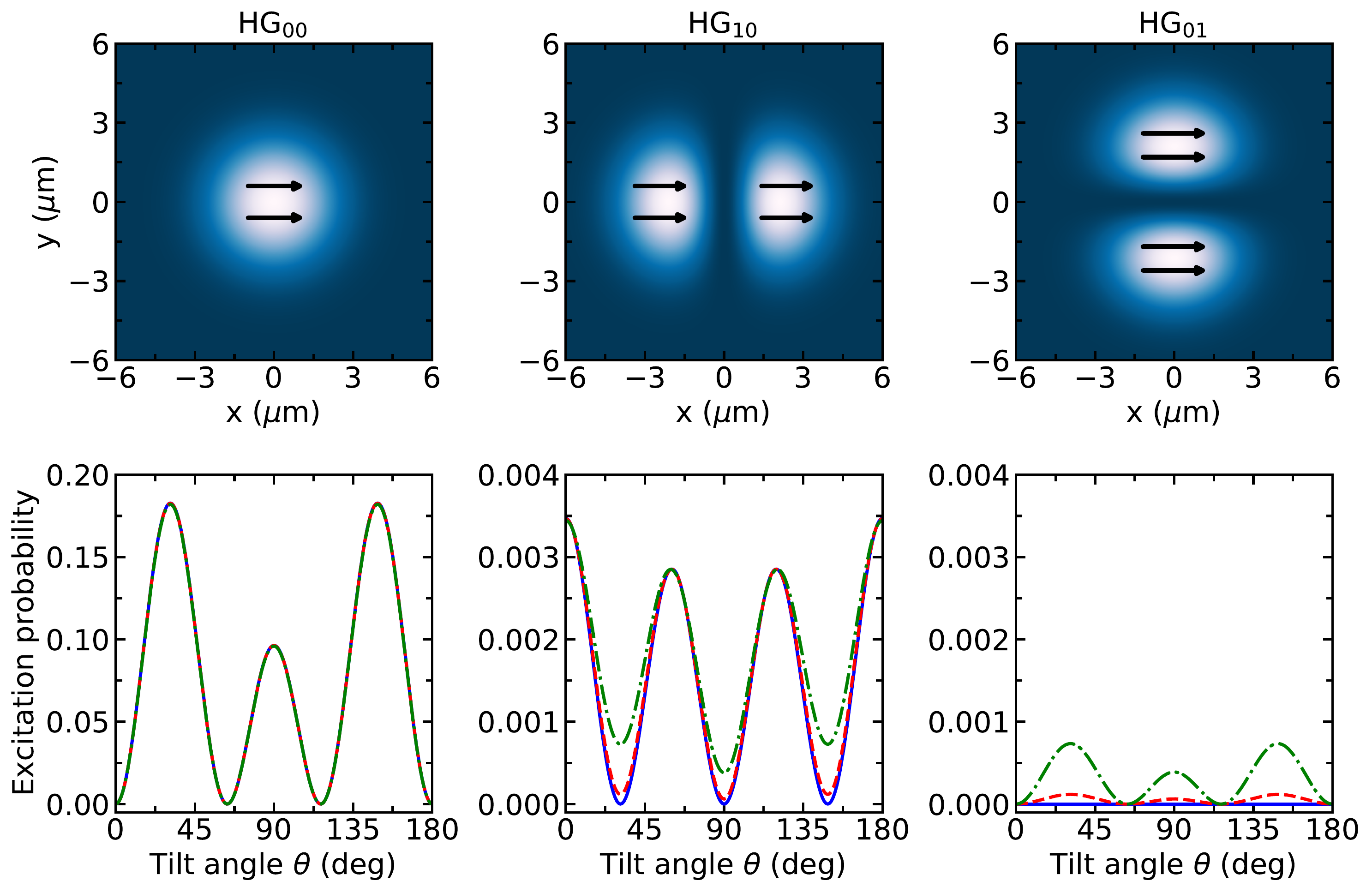}
	\caption{Top: Intensity profiles for Hermite–Gaussian modes HG$_{00}$ (left), HG$_{10}$ (middle), and HG$_{01}$ (right), linearly polarized along the $x$-axis. Bottom: Corresponding probabilities of the $|F_g \! = \! 0, \, M_g \! = \! 0 \rangle + \gamma \to \, |F_e \! = \! 3, \, M_e \! = \! 0 \rangle$ transition as a function of the magnetic field tilt angle $\theta$ for a single $^{171}$Yb$^{+}$ ion placed exactly at the beam center (blue solid lines) or exhibiting spatial distribution with width parameter $\sigma = 40$ nm (red dashed lines) and $\sigma = 100$ nm (green dash-dotted lines). Calculations were performed for the wavelength of 467 nm, the beam waist $w_0 = 3$ $\mu$m, the total power $P = 1$ mW, and the pulse duration $t = 1.5$ ms.}
	\label{fig:probX}
\end{figure*}
\subsection{Hermite-Gaussian modes}
\label{sec:hg}
Having discussed how to compute the transition amplitudes for Bessel (and LG) radiation, we can consider now photoexcitation of an atom by Hermite-Gaussian modes HG$_{mn}$. These are known to be solutions of the paraxial wave equation in Cartesian coordinates and are usually characterized by two indices $m$ and $n$, which determine their transverse intensity profile \cite{Andrews:2013}. In our study we will focus on the modes HG$_{00}$ (Gaussian beam), HG$_{10}$, and HG$_{01}$ that are planned to be used in future clock on a chip experiments at PTB.

In contrast to diffraction-free Bessel solutions, the intensity distribution of which is the same in every plane normal to the beam axis, the width of HG intensity profile changes during propagation. In planned experiments, a target atom or ion will be placed in the plane of minimum width, at $z=0$. In this plane the vector potentials of the HG$_{00}$, HG$_{10}$, and HG$_{01}$ modes with the frequency $\omega$ and the beam waist $w_0$ can be written as:
\begin{subequations}
\label{eq:hg}
\begin{align}
	\bm{A}^{(\text{HG}_{00})} (x, y) &= \bm{e} A_0 \, e^{-(x^2+y^2)/w_0^2}  \, , \\
	\bm{A}^{(\text{HG}_{10})} (x, y) &= \bm{e} A_0 \, \frac{2x}{w_0} \, e^{-(x^2+y^2)/w_0^2}  \, , \\
	\bm{A}^{(\text{HG}_{01})} (x, y) &= \bm{e} A_0 \, \frac{2y}{w_0} \, e^{-(x^2+y^2)/w_0^2}  \, ,
\end{align}
\end{subequations}
where $\bm{e}$ is the polarization vector \cite{Gbur:2017}. In Eqs.~\eqref{eq:hg}, the constant $A_0 = \sqrt{4P/(c \epsilon_0 \omega^2 \pi w_0^2)}$ is chosen in such a way that an integral of the intensity $I = c \epsilon_0 \omega^2 |\bm{A}|^2 /2$ over the beam cross-section gives the total power $P$ \cite{Fox:2006, Akhmanov:1997}. It follows from Eqs.~\eqref{eq:hg} that the HG$_{10}$ and HG$_{01}$ modes have one nodal line directed along $y$ and $x$ axes, respectively. As seen from the upper panel of Fig.~\ref{fig:probX}, each of these lines separates two bright spots. Thus, in contrast to the case of Bessel and Gaussian modes, the HG$_{10}$ and HG$_{01}$ solutions do not have axial symmetry with respect to the direction of light propagation.

\begin{figure*}[t]
    \centering
	\includegraphics[width=0.76\linewidth]{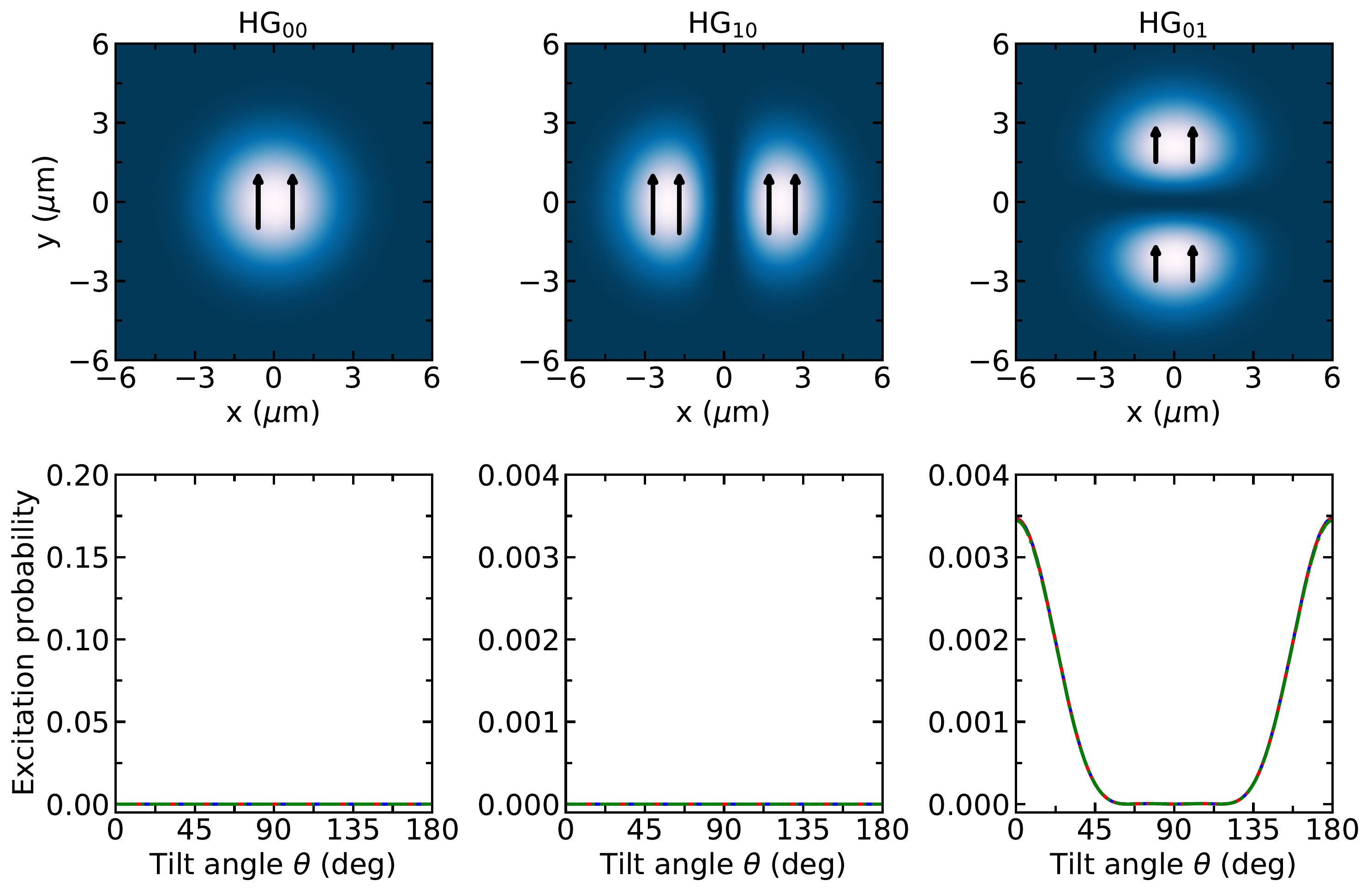}
	\caption{Same as Fig.~\ref{fig:probX}, but for the linear polarization along the $y$-axis.}
	\label{fig:probY}
\end{figure*}

Rigorous calculation of the transition amplitudes for Hermite-Gaussian modes is somewhat more complicated than for the case of Bessel radiation. Fortunately, in the region near the beam center located at $x=y=0$ the HG$_{00}$, HG$_{10}$, and HG$_{01}$ modes \eqref{eq:hg} can be
well approximated by a linear combination of Bessel solutions \eqref{eq:bes} in the paraxial regime. In particular, for the circularly polarized Hermite-Gaussian beams, we find:
\begin{subequations}
\label{eq:hg_bc}
\begin{align}
	\bm{A}^{(\text{HG}_{00})}_{\lambda = \pm 1} &\approx  \pm i A_0 \, \bm{A}^{(\text{B, par})}_{m_l = 0, \, \lambda = \pm 1} \, , \\
	\bm{A}^{(\text{HG}_{10})}_{\lambda = \pm 1} &\approx  \pm 1.04 \, \frac{i A_0}{\sqrt{2}} \left[ \bm{A}^{(\text{B, par})}_{m_l = +1, \, \lambda = \pm 1} - \bm{A}^{(\text{B, par})}_{m_l = -1, \, \lambda = \pm 1} \right] \, , \\
	\bm{A}^{(\text{HG}_{01})}_{\lambda = \pm 1} &\approx  \pm 1.04 \, \frac{A_0}{\sqrt{2}} \left[ \bm{A}^{(\text{B, par})}_{m_l = +1, \, \lambda = \pm 1} + \bm{A}^{(\text{B, par})}_{m_l = -1, \, \lambda = \pm 1} \right] \, ,
\end{align}
\end{subequations}
where 
\begin{align}
	\label{eq:bes_par}
	\bm{A}_{m_l \lambda}^{(\text{B, par})}  (\bm{r}) &\approx \bm{A}_{m_{\gamma} \lambda}^{(\text{B})}  (\bm{r}; \, \varkappa \ll k_z) \, \notag \\
    &= \bm{e}_{\lambda} \, (-i)^{\lambda} J_{m_l}(\varkappa r_{\perp}) \, e^{i m_l \phi} \, e^{i k_z z} \, .
\end{align}
The latter expression is obtained from Eq.~\eqref{eq:bes} by integrating over $\bm{k}_{\perp}$ under the paraxial condition $\varkappa \ll k_z$, see Ref.~\cite{Schulz/PRA:2020} for further details. In Eq.~\eqref{eq:bes_par}, $m_l = m_\gamma - \lambda$ denotes the projection of the orbital angular momentum, and $\bm{e}_{\lambda} \equiv \bm{e}_{\bm{k} \parallel z, \lambda}$ with $\lambda = \pm 1$ stands for the vector describing the states of right-hand and left-hand circularly polarized radiation.

In order to approximate the HG solutions \eqref{eq:hg} by the paraxial Bessel beams \eqref{eq:bes_par}, we have also adjusted the ratio of the transverse to the longitudinal photon momentum components determined by the opening angle $\theta_k$, as well as the general prefactors. In particular, while for HG$_{00}$ we have chosen $\theta_k = \arcsin [2 / (w_0 k)]$, the opening angle $\theta_k = \arcsin [2.6 / (w_0 k)]$ and the prefactor 1.04 are taken for HG$_{10}$ and HG$_{01}$. For these choice of parameters, Eqs.~\eqref{eq:hg} and \eqref{eq:hg_bc} give almost equivalent vector potentials in the vicinity of the beam center located at $x=y=0$. For example, for the case of the modes with wavelength 467 nm and beam waist 3 $\mu$m displayed in the upper panel of Fig.~\ref{fig:probX}, this region is bounded by $-2$ $\mu$m $<$ $x$ $<$ $2$ $\mu$m and $-2$ $\mu$m $<$ $y$ $<$ $2$ $\mu$m. This is sufficient to describe future experiments with Doppler cooled atoms in which atom localization near the beam center below 40 nm is expected due to thermal fluctuations. For the chosen parameters, we can also neglect the effects of the $z$-dependence of the beam intensity profile, since its Rayleigh range of about 60 $\mu$m (i.e.~the distance at which the beam width increases noticeably) is much larger than the expected fluctuations of the $z$ coordinate of an atom.

By making use of Eqs.~\eqref{eq:hg_bc}, we can express the transition amplitudes for excitation of an atom by circularly polarized HG modes
\begin{subequations}
\label{eq:melement_hg_c}
\begin{align}
	\mathcal{M}_{M_e M_g}^{(\text{HG}_{00})} (\lambda = \pm 1) &\approx \pm i \, A_0 \, \mathcal{M}_{M_e M_g}^{(\text{B, par})} (m_l = 0, \, \lambda = \pm 1) \, , \\
	\mathcal{M}_{M_e M_g}^{(\text{HG}_{10})} (\lambda = \pm 1) &\approx  \pm 1.04 A_0 \, \frac{i}{\sqrt{2}} \notag \\ 
	&\times \left[ \mathcal{M}_{M_e M_g}^{(\text{B, par})} (m_l = +1, \, \lambda = \pm 1) \right. \notag \\
	&\left. \;\; - \mathcal{M}_{M_e M_g}^{(\text{B, par})} (m_l = -1, \, \lambda = \pm 1) \right] \, , \\
	\mathcal{M}_{M_e M_g}^{(\text{HG}_{01})} (\lambda = \pm 1) &\approx  \pm 1.04 A_0 \, \frac{1}{\sqrt{2}} \notag \\
	&\times \left[ \mathcal{M}_{M_e M_g}^{(\text{B, par})} (m_l = +1, \, \lambda = \pm 1) \right. \notag \\
	&\left. \;\; + \mathcal{M}_{M_e M_g}^{(\text{B, par})} (m_l = -1, \, \lambda = \pm 1) \right] \, ,
\end{align}
\end{subequations}
in terms of their Bessel counterparts \eqref{eq:melement2}. These formulas can be readily used to obtain the amplitudes for linearly polarized light: 
\begin{subequations}
\label{eq:melement_hg_xy}
\begin{align}
	\mathcal{M}_{M_e M_g}^{(\text{HG}_{00, 10, 01})} (x) =&  \frac{1}{\sqrt{2}} \left[ \mathcal{M}_{M_e M_g}^{(\text{HG}_{00, 10, 01})} (\lambda = +1) \right. \notag \\
	&\left. \;\;\;\;+ \mathcal{M}_{M_e M_g}^{(\text{HG}_{00, 10, 01})} (\lambda = -1) \right] \, , \\
	\mathcal{M}_{M_e M_g}^{(\text{HG}_{00, 10, 01})} (y) =&  \frac{i}{\sqrt{2}} \left[ \mathcal{M}_{M_e M_g}^{(\text{HG}_{00, 10, 01})} (\lambda = -1) \right. \notag \\
	&\left. \;\;\;\;- \mathcal{M}_{M_e M_g}^{(\text{HG}_{00, 10, 01})} (\lambda = +1) \right] \, ,
\end{align}
\end{subequations}
where we have employed the standard relationships $\bm{e}_x = (\bm{e}_{\lambda=+1} + \bm{e}_{\lambda=-1})/\sqrt{2}$ and $\bm{e}_y = i(\bm{e}_{\lambda=-1} - \bm{e}_{\lambda=+1})/\sqrt{2}$ between linear and circular polarization unit vectors \cite{Rose:1957}.

%
%
\section{Results and discussion}
\label{sec:results}
While the theory presented above can be applied to describe excitation of an arbitrary atom by low-order HG modes, here we will consider the $4f^{14} 6s \; ^2S_{1/2} (F \! = \! 0) \to \; 4f^{13} 6s^2 \; ^2F_{7/2} (F \! = \! 3)$ transition in the $^{171}$Yb$^{+}$ ion. This 467 nm transition, proceeding via the E3 channel, is of great interest since it serves as the reference for high-precision optical clocks and enables searches for new physics \cite{Dzuba/NP:2016, Sanner/N:2019, Dreissen/NC:2022}. Below we shall focus our attention on the transition between the magnetic sublevels $M_g = M_e = 0$ that can be spectroscopically resolved by the Zeeman effect. As seen from Eq.~\eqref{eq:melement2}, the evaluation of the transition amplitudes requires a knowledge of the reduced matrix element $\langle 4f^{13} 6s^2 \, ^2F_{7/2} || H_{\gamma} (E3) || 4f^{14} 6s \, ^2S_{1/2} \rangle$. Its calculation for $^{171}$Yb$^{+}$ is a very complicated task, requiring the use of sophisticated atomic structure theories. To avoid this, we can estimate the matrix element from the expression
\begin{align}
	\label{eq:reduced}
	\langle 4f^{13} 6s^2 \, ^2F_{7/2} || H_{\gamma} (E3) || 4f^{14} 6s \, ^2S_{1/2} \rangle = \sqrt{\frac{1}{\pi \alpha \omega \tau}}  \, ,
\end{align}
which relates it to the measured lifetime $\tau = 4.98 \times 10^7$ s of the $^2F_{7/2}$ excited state \cite{Lange/PRL:2021, Popov:2017}. We shall also assume that the total power of the beam is $P = 1$ mW and its waist is $w_0 = 3$ $\mu$m. A similar parameter range has been employed in recent experiments on excitation of the E3 transition in $^{171}$Yb$^{+}$ ion by LG light \cite{Lange/PRL:2022}.

\subsection{Transition probability}
\label{sec:ex}
We are ready now to use Eqs.~\eqref{eq:melement_hg_xy} to compute the probability $W$ of excitation of an $^{171}$Yb$^{+}$ ion by linearly polarized HG modes. For a well-defined impact parameter of an atom, this transition probability is given by $W (\bm{b}) = \left|ec \mathcal{M}^{(\text{HG})} (\bm{b}) /\hbar \right|^2 t^2/4$. This approximate formula was derived from the analysis of Rabi oscillations for short interaction times $t$ \cite{Auzinsh:2010}. In calculations below we assume $t = 1.5$ ms which is much smaller than the inverse Rabi frequency. 

As already mentioned above, in real experiments it is not possible to achieve perfect control of the atom's position. To account for such uncertainty, we assume that the impact parameter $\bm{b}$ follows a Gaussian distribution with a width $\sigma$:
\begin{align}
	\label{eq:distr}
	f (\bm{b}) = \frac{1}{2 \pi \sigma^2} e^{-\frac{\bm{b}^2}{2 \sigma^2}} \, .
\end{align}
With the help of $f (\bm{b})$, we can use a semi-classical approximation to express the transition probability for a single atom target, centered on the beam axis, as: 
\begin{align}
	\label{eq:average}
	W^{(\text{HG})}_{x, \,y} = \frac{(ect)^2}{4 \hbar^2} \int f (\bm{b}) \, \left| \mathcal{M}_{M_e M_g}^{(\text{HG})} \left( x, y \right) \right|^2 \, d^2 \bm{b} \, .
\end{align}
In the past this approximation has been successfully employed to describe experimental results obtained for LG modes \cite{Lange/PRL:2022}.  

Fig.~\ref{fig:probX} shows the transition probabilities as a function of the magnetic field tilt angle $\theta$ for HG$_{00}$ (left), HG$_{10}$ (middle), and HG$_{01}$ (right) modes, linearly polarized along the $x$-direction. Here, the blue solid line represents the results for an atom placed exactly at the beam center ($b=0$, $\sigma=0$), while the red dashed and green dash-dotted lines correspond to the predictions of Eq.~\eqref{eq:average} for the spatial distribution $f (\bm{b})$ with width parameters $\sigma = 40$ nm and $100$ nm, respectively. Since beam-pointing fluctuations are expected to be negligible in an integrated optics setup, we estimated these parameters based on the thermal spread of an spatial atomic wavepacket. In this case, $\sigma$ can be derived from the relation $m_\text{a} \omega_\text{r}^2 \sigma^2 /2 = k_\text{B} T /2$ for a classical harmonic oscillator, where $m_\text{a}$ is the mass of an ion and $T$ is its temperature. With the help of this expression, we obtained the above mentioned widths $\sigma$ for a realistic trapping frequency $\omega_\text{r} = 2\pi \times 600$ kHz and temperatures $T = T_\text{Doppler} = 0.5$ mK corresponding to the Doppler cooling limit and $T = 3.1$ mK. In the later discussion of the light shift we will also consider a smaller width $\sigma = 10$ nm, which corresponds to the size of the quantum mechanical ground-state wave function of the trapped Yb$^{+}$ ion. We see that the excitation probability $W^{(\text{HG}_{00})}_{x}$ for the Gaussian beam is insensitive to variations of the target size. This can be expected since the Gaussian beam size $\approx 3$ $\mu$m is large compared to any of the width parameters considered here. Therefore, the intensity and phase of HG$_{00}$ mode are almost constant over the entire spread of a target. For this case, the excitation probability exhibits the well-known plane-wave behavior \cite{Schulz/PRA:2020}, reaching maximum values at angles $\theta = 31^{\circ}$, $90^{\circ}$, and $149^{\circ}$.

A different behavior of $W^{(\text{HG})}_{x}$ can be observed for HG$_{10}$ and HG$_{01}$ modes. For these two cases, the delocalization of an atom may influence the excitation probability; however, the effect becomes pronounced only for very large targets, $\sigma \approx 100$ nm. Such a $\sigma$-dependence is caused by the inhomogeneity of the electric field distribution of HG$_{10}$ and HG$_{01}$ beams. Moreover, their complex internal structure also affects the $\theta$-dependence of the transition probability. For example, $W^{(\text{HG}_{10})}_{x}$ is
enhanced at $\theta = 0^{\circ}$, $60^{\circ}$, $120^{\circ}$, and $180^{\circ}$, which is much different from what was observed for the Gaussian regime. In contrast, $W^{(\text{HG}_{01})}_{x}$ exhibits a qualitatively similar $\theta$-dependence as $W^{(\text{HG}_{00})}_{x}$, but is strongly suppressed and becomes observable only for large targets. 

\begin{figure}[t!]
    \centering
	\includegraphics[width=0.96\linewidth]{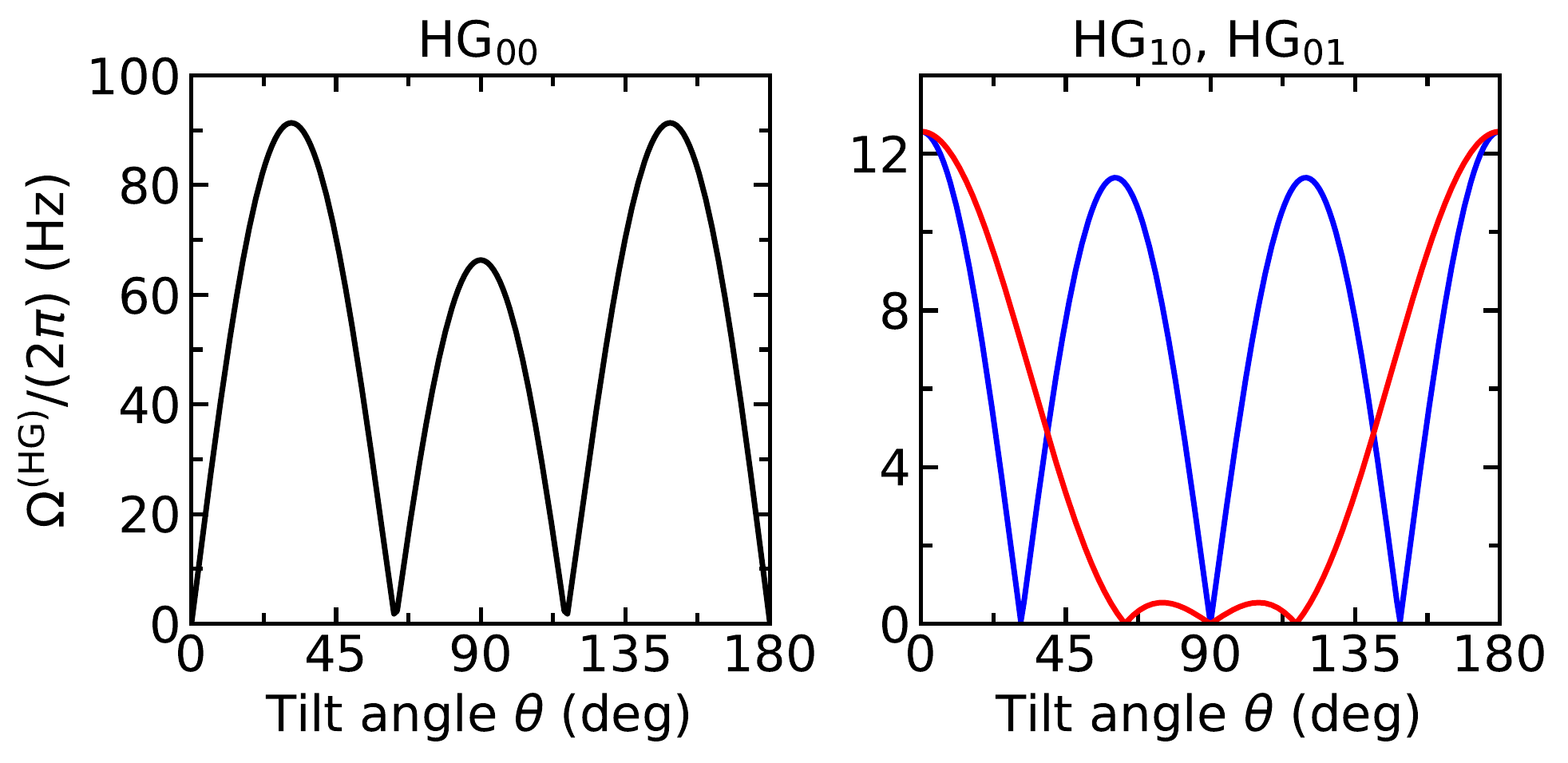}
	\caption{Rabi frequencies \eqref{eq:rabi2} as a function of the magnetic field tilt angle $\theta$ for HG$_{00}$ (black line) and HG$_{10}$ (blue line) modes linearly polarized in the $x$-direction and for HG$_{01}$ (red line) mode linearly polarized in the $y$-direction. The atom is placed in the beam center ($b=0$, $\sigma=0$). All other parameters are the same as in Fig.~\ref{fig:probX}.}
	\label{fig:rabi}
\end{figure}

Fig.~\ref{fig:probY} displays the excitation probabilities for the same three modes, HG$_{00}$, HG$_{10}$, and HG$_{01}$, but for linear polarization in the $y$-direction. As seen from the figure, HG$_{00}$ and HG$_{10}$ light beams hardly induce the $M_g \! = \! 0 \to \,  M_e \! = \! 0$ transition, regardless of the tilt angle $\theta$ of the magnetic field. This suppression of the E3 transition can be explained using symmetry arguments. For example, for plane-wave radiation and the transition of interest, a detailed discussion of zero excitation probability based on the symmetry analysis of ionic states and of incident light has been presented in Ref.~\cite{Schulz/PRA:2020}. Such an analysis can be extended to HG$_{00}$ and HG$_{10}$ modes when an atom is placed on the vortex line, $b=0$, but is not fully justified for a delocalized atom. Nevertheless, for the latter case, $\sigma > 0$, the probability of the E3 transition is very tiny and hence is not seen in the left and middle panels of Fig.~\ref{fig:probY}. In contrast, excitation with HG$_{01}$ is possible for small and large tilt angles and is most pronounced for $\theta = 0^{\circ}$ and $180^{\circ}$.

\subsection{Rabi frequency and light shift}
\label{sec:rabi}
As seen from Figs.~\ref{fig:probX} and \ref{fig:probY}, the results obtained for a perfectly localized atom, $b=0$ and $\sigma=0$, differ only little from the predictions of Eq.~\eqref{eq:average} for $\sigma = 40$ nm at Doppler temperature. We argue therefore that the atomic spread effects can be neglected in the analysis of future experiments in which the size of a target is much smaller than the beam waist. As mentioned above, this is the case for planned studies at PTB, where an integrated optics scheme improves the pointing stability of the laser beam relative to the ion. The approximation of a perfectly localized atom allows one to derive simple analytical expressions for the Rabi frequency:
\begin{align}
    \label{eq:rabi1}
	\Omega^{(\text{HG})} = \frac{ec}{\hbar} \left|  \mathcal{M}_{M_e M_g}^{(\text{HG})} (b=0) \right| \, .
\end{align}
Indeed, by making use of Eqs.~\eqref{eq:melement_hg_xy} and \eqref{eq:rabi1} for the case of the $|F_g \! = \! 0, \, M_g \! = \! 0 \rangle + \gamma \to \, |F_e \! = \! 3, \, M_e \! = \! 0 \rangle$ $E3$ transition, induced by various linearly polarized Hermite-Gaussian modes, we find:
\begin{subequations}
\label{eq:rabi2}
\begin{align}
	\Omega^{(\text{HG}_{00})}_{x} &\approx \sqrt{\frac{6P}{\hbar \omega k^2 w_0^2 \tau}} \, \left| \sin \theta (5\cos^2 \theta - 1) \right| \, , \\
	\Omega^{(\text{HG}_{10})}_{x} &\approx 2.7 \sqrt{\frac{3P}{\hbar \omega k^4 w_0^4 \tau}} \, \left| \cos \theta (15\cos^2 \theta - 11) \right| \, , \\
	\Omega^{(\text{HG}_{01})}_{y} &\approx 2.7 \sqrt{\frac{3P}{\hbar \omega k^4 w_0^4 \tau}} \, \left| \cos \theta (5\cos^2 \theta - 1) \right| \, , \\
    \Omega^{(\text{HG}_{00})}_{y} &= \Omega^{(\text{HG}_{10})}_{y} = \Omega^{(\text{HG}_{01})}_{x} \approx 0 \, ,
\end{align}
\end{subequations}
where we have employed the paraxial approximation of the Bessel amplitude obtained from Eq.~\eqref{eq:melement2} for small opening angles $\theta_k$.

In Fig.~\ref{fig:rabi} we show non-zero Rabi frequencies $\Omega^{(\text{HG}_{00})}_{x}$, $\Omega^{(\text{HG}_{10})}_{x}$, and $\Omega^{(\text{HG}_{01})}_{y}$, calculated for the set of parameters mentioned at the beginning of Sec.~\ref{sec:results}. We see that while the Rabi frequency may reach $\Omega^{(\text{HG}_{00})}_{x} \approx 2\pi \times 90$ Hz for the Gaussian mode and $\theta = 31^{\circ}$ or $149^{\circ}$, its maximum values are $\Omega^{(\text{HG}_{10})}_{x} = \Omega^{(\text{HG}_{01})}_{y} \approx 2\pi \times 12$ Hz for the HG$_{10}$ and HG$_{01}$ light beams and $\bm{B} \parallel \bm{k}_z$. It is also informative to compare $\Omega^{(\text{HG}_{10})}$ and $\Omega^{(\text{HG}_{01})}$ with the corresponding Rabi frequencies for Laguerre-Gaussian modes that were used in the previous experiments \cite{Lange/PRL:2022}. By using Eq.~\eqref{eq:bes_par} and the theory from above, we find that $\Omega^{(\text{HG}_{10})}_{x} / \Omega^{(\text{LG}_{01})}_{x} = \Omega^{(\text{HG}_{01})}_{y} / \Omega^{(\text{LG}_{01})}_{y} = \sqrt{2}$, thus indicating that excitation probability of the E3 transition induced by HG modes is two times larger than that obtained by LG light. Moreover, we estimated for our beam parameters and an optimal excitation geometry that an $^{171}$Yb$^{+}$ ion placed in the center of HG$_{10}$ or HG$_{01}$ modes experiences the light shift in the range from $\Delta \nu = 50$ Hz to 8 Hz if the target size varies from $\sigma = 100$ nm to 40 nm \cite{Lange/PRL:2021, Itano/JRNIST:2000}. An even smaller light shift of about $\Delta \nu = 0.5$ Hz can be achieved for the ``quantum-mechanical'' width $\sigma = 10$ nm. These values are comparable to light shifts for LG$_{01}$ modes and are much smaller than $\Delta \nu = 205$ Hz obtained for a Gaussian beam with power adjusted to provide a similar Rabi frequency.

%
%
%
\section{Conclusions and outlook}
\label{sec:summary}

In this paper we propose using Hermite-Gaussian modes in photonic integrated circuits to induce dipole-forbidden atomic clock transitions. In order to investigate the feasibility of this scheme, we present a theoretical analysis of the excitation of a single trapped atom by HG modes. By expressing the HG vector potentials in terms of their Bessel counterparts, written in the paraxial approximation, we were able to derive simple analytical expressions for the Rabi frequencies $\Omega^{\rm (HG)}$ of the photoinduced transitions. These expressions allow one to analyze the dependence of $\Omega^{\rm (HG)}$ both on the parameters of HG beams and on the orientation of the external magnetic field, used to define the quantization axis of a target ion. The developed theory was applied to investigate the case of the $^{2}S_{1/2} \to \; ^{2}F_{7/2}$ E3 transition in an $^{171}$Yb$^{+}$ ion localized in the beam's center. Calculations, performed for experimentally feasible parameters, helped us to determine optimal combinations of incident light polarization and orientation of the magnetic field, resulting in rather large Rabi frequencies. For HG$_{10}$ and HG$_{01}$ modes, we have estimated that $\Omega^{\rm (HG)}$ lies in the range of tens of Hz, which can exceed $\Omega^{(\text{LG})}$ of the previously used ``standard'' Laguerre-Gaussian light for the same beam waist and power. This suggests that Hermite-Gaussian beams can serve as a useful tool for studying highly-forbidden atomic transitions. 

In the present theoretical study we have used a ``semiclassical'' approach, Eq.~\eqref{eq:average}, to account for the delocalization of a target ion with respect to the beam center. Even though this approach is expected to provide reliable qualitative predictions for the dependence of the Rabi frequency on the geometry of the experiment and on the state of the incident HG beam, it should be probed against the fully quantum theory in which the center-of-mass motion of a target atom is quantized. Similar studies have been performed for an atomic target exposed to standing electromagnetic waves \cite{Vasquez/PRL:2023}. The fully quantum analysis employing the time-dependent density matrix approach \cite{Peshkov/PRA:2023} is currently underway and will be presented in a follow-up publication.

Following our theoretical analysis we propose a photonic integrated circuit (PIC) frequency reference based on an octupole transition in Yb$^+$ ions that can be used in a PIC-based optical clock. Such a clock will be orders of magnitude more compact compared to existing setups and have lower power consumption and higher pointing stability, making it well suited for applications outside the laboratory, including deep space navigation and geodetic measurements. As a first experiment in this direction, we plan to use $x$-polarized HG$_{10}$ or $y$-polarized HG$_{01}$ modes in a ``clock-on-a-chip'' setup. For lowest light shifts, we plan to operate the clock with $^{173}$Yb$^+$ which offers 10 times larger transition matrix elements and thus a 100 fold reduction of the light shift on its clock transitions \cite{Lange/PRL:2021, Dzuba/PRA:2016}. Since the linewidth of the atomic transition is still narrower than the laser linewidth, depending on the laser noise spectrum the suppression factor can be even larger.

\section*{Acknowledgments}
This work was funded by the Deutsche Forschungsgemeinschaft (DFG, German Research Foundation) -- Project-ID 445408588 (SU 658/5-1) and under Germany's Excellence Strategy - EXC-2123 QuantumFrontiers - 390837967. This work was also supported by the Max-Planck-RIKEN-PTB-Center for Time, Constants and Fundamental Symmetries. We thank C.-F.~Grimpe for the discussion about integrated optics.

%
%

\end{document}